\documentclass[aps, pre, twocolumn, a4paper, superscriptaddress, floatfix]{revtex4}
\usepackage{graphicx}
\usepackage{amsmath}
\usepackage{placeins}
\usepackage{color}
\usepackage{hyperref}

\newcommand{\beq}{\begin{equation}}
\newcommand{\eeq}{\end{equation}}
\newcommand{\beqa}{\begin{eqnarray}}
\newcommand{\eeqa}{\end{eqnarray}}

\begin{document}

\title{Emergence of stylized facts during the opening of stock markets}
\author{Sebastian M. Krause}
\author{Jonas A. Fiegen}
\author{Thomas Guhr}
\affiliation{Faculty of Physics, University of Duisburg-Essen, Duisburg, Germany}

\begin{abstract}
Financial markets show a number of non-stationarities, ranging from volatility fluctuations over ever changing technical and regulatory market conditions to seasonalities. On the other hand, financial markets show various stylized facts which are remarkably stable. It is thus an intriguing question to find out how these stylized facts emerge. As a first example, we here investigate how the bid-ask-spread between best sell and best buy offer for stocks develops during the trading day. For rescaled and properly smoothed data we observe collapsing curves for many different NASDAQ stocks, with a slow power law decline of the spread during the whole trading day. This effect emerges robustly after a highly fluctuating opening period. Some so called large-tick stocks behave differently because of technical boundaries. Their spread closes to one tick shortly after the market opening. We use our findings for identifying the duration of the market opening which we find to vary largely from stock to stock. 
\end{abstract}

\maketitle

\section{Introduction}

Financial markets are highly non-stationary~\cite{mandelbrot1997variation,mantegna1997stock,munnix2012marketstates}. Market conditions change constantly due to technical innovations and regulatory changes~\cite{chung2016high}. The correlations between stock price returns of different stocks change over time ~\cite{rosenow2003dynamics,munnix2012marketstates}. Correlations are especially strong in times of crisis and correlated losses imply systemic risk. Further the volatility of stock prices varies strongly, with small volatility during calm periods and large volatility during times of crisis~\cite{cont2007volatility}. Even though system properties as volatility vary strongly over time, they can be described with universal scaling laws~\cite{mantegna1995scaling,gabaix2003theory,bouchaud2001power}, establishing different financial stylized facts as the inverse cubic power-law distribution of returns~\cite{plerou1999scaling,drozdz2007stock} and slowly decaying auto-correlation of absolute returns caused by volatility clustering~\cite{liu1999statistical,cont2007volatility}. Stylized facts apply robustly even for markets under drastically changing conditions from early industrialization to highly digitalized markets. 

Another sign of non-stationarity in financial markets is seasonality. The time of the year~\cite{rozeff1976capital,chang2016being}, official reports~\cite{chang2016being,francis1992stock}, holidays~\cite{ariel1990high}, the day of the week~\cite{brooks2001seasonality} and the time of the day influence the market dynamics. Particularly strong is intra-day seasonality, with pronounced over-night effects~\cite{francis1992stock}, a special character of the market opening period~\cite{brock1992periodic}, and frozen dynamics of the price after the market closes~\cite{clara2017diffusive}. 
Intra-day volatility was intensely investigated~\cite{bauwens2013econometric}. Possible reasons for seasonalities were discussed, such as arrival of new information over night or during the day, habits of traders, and the need to close positions before the market closes. 
A profound understanding of intra-day seasonality is important for modeling volatility on longer time scales
~\cite{genccay2001differentiating}. Seasonality-adjusted agent based modeling~\cite{giardina2001microscopic,cont2007volatility,krause2013spin,krause2012opinion} of the order book mechanics can shed light on mini flash-crashes~\cite{filimonov2012quantifying,braun2018impact} and seasonal impact effects~\cite{webb2016price}. 

A fundamental open question is how stylized facts emerge in highly non-stationary financial markets. Here we investigate as a first example the intra-day seasonality of the bid-ask-spread of stocks. The spread is the price difference between best sell and best buy offer. It is an important measure for market liquidity and trading costs, and its average value over the trading day strongly impacts the orderflow and the structure of the order book~\cite{gareche2013fokker,dayri2015large,theissen2017regularities}. We analyze order flow data of $96$ NASDAQ stocks in early 2016. The spread performs large fluctuations during the whole trading day, and there are large differences  between different stocks concerning activity and sensitivity to technical bounds as the tick size~\cite{theissen2017regularities}. To overcome these problems we design a robust method for calculating moving averages which adapts to individual activity and smooths fluctuations. This allows us to uncover how the market organizes to a non-stationary state with universal time dependence of the spread, instead of approaching a constant stationary-state value. Only if the spread reaches the lower technical bound of one tick it converges to this value. In both cases we observe an opening period with different behavior compared to the rest of the trading day and also with larger fluctuations. We quantify the duration of this opening period individually for every stock.

\section{Data}\label{sec:data}

We analyze Historical TotalView-ITCH data from NASDAQ US, downloaded from \cite{tp}. 
Our data span over five consecutive working days from March, 7 2016 to March, 11 2016. Out of the 100 stocks listed in the NASDAQ 100 in this period \cite{wikiNasdaq100}, four stocks are not available in \cite{tp}, and therefore cannot be included in our analysis. 
The data provide information about limit orders being placed,
deleted, partially canceled, partially traded and fully traded.
Moreover, they contain information about trades against hidden orders. A detailed description of the data can be found in \cite{huang2011lobster}.

We analyze data from times between the market opening time $t_0=$ 9:30 am and the market closing $t_e=$ 4:00 pm (New York time). 
All events have a time stamp in milliseconds.
Events happening in the same millisecond have the same time stamp. Incoming orders are processed by the market in the same order as in which they arrive, even if the time stamp is the same.

For every stock and day individually we reconstruct the orderbook of all visible orders. This allows us to identify for every time $t$ the best available ask price bestask$(t)$, and the best available bid price bestbid$(t)$. For characterizing the price of the stock irrespective of selling or buying, the midpoint price 
\begin{align}
m(t) &= \frac{{\rm bestbid(t)}+{\rm bestask(t)}}{2}
\end{align}
is a good measure. The difference between best bid and best ask price is the spread
\begin{align}
s(t) &= {\rm bestask}(t)-{\rm bestbid}(t).
\end{align}
This is an important measure for the market liquidity. If the spread is small, trading is possible with low costs. 
The structure of the orderbook beyond the quotes of best bid and best ask depends strongly on the spread $s$~\cite{theissen2017regularities}. If the spread is as small as one tick, a stock is characterized as large-tick stock~\cite{eisler2012price,gareche2013fokker,dayri2015large}. In this case, the      quotes carry large volumes, the orderbook close to the quotes is densely filled, but far from the quotes the orderbook is sparse. A spread size of many ticks usually results in gaps behind the quotes, but limit orders far from the quotes are more likely. As the spread $s$ is an important characteristic of the whole orderbook, we will concentrate on this quantity in the following. 

\section{Emergence of scaling}\label{sec:scaling}

\begin{figure}[htb]
\begin{center}
    \includegraphics[width=1.0\columnwidth]{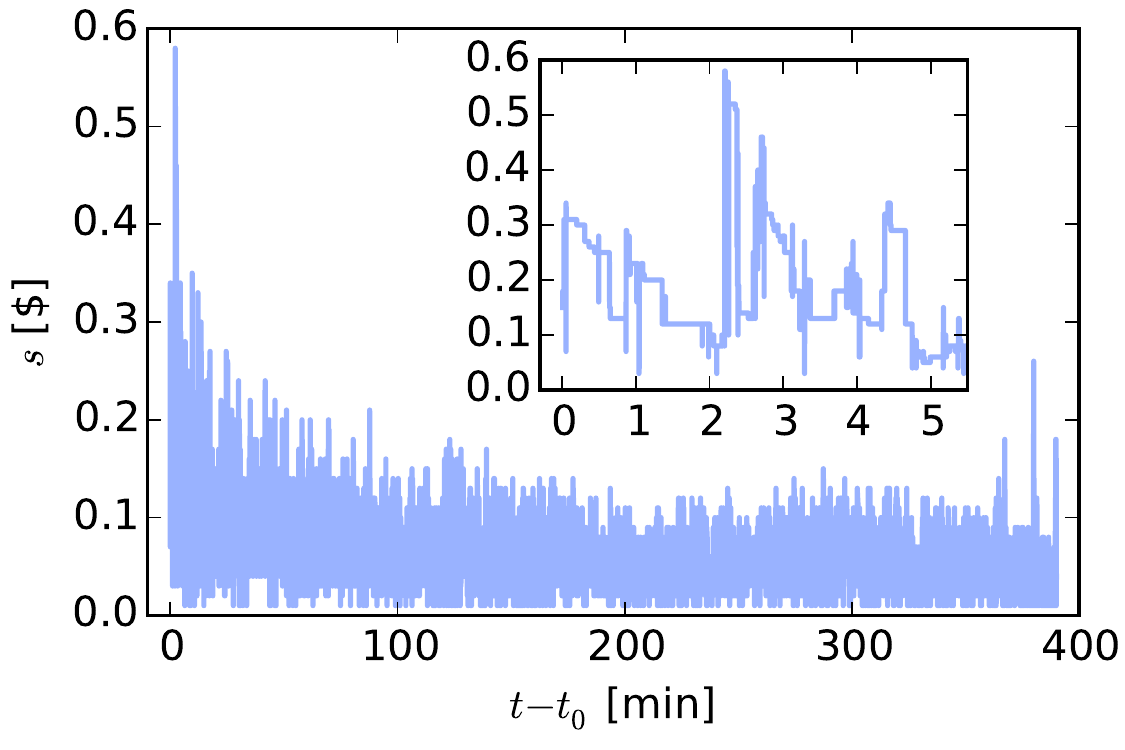}
    \includegraphics[width=1.0\columnwidth]{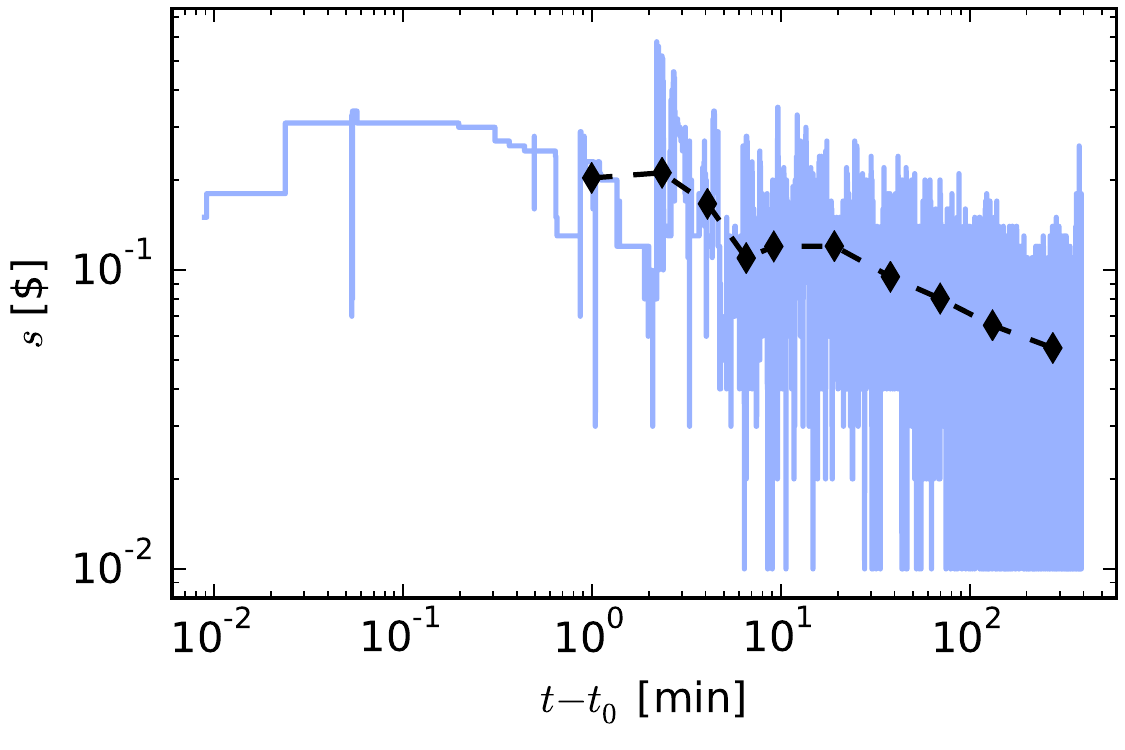}
    \caption{Blue curves show the spread of Celgene Corporation on Monday. The inset in the upper panel is a zoom into the first minutes. The lower panel shows the same data on double logarithmic scales, with black diamonds representing the moving average.}
    \label{fig:spread_CELG}
\end{center}
\end{figure}
In the upper panel of Fig.~\ref{fig:spread_CELG} the spread $s(t-t_0)$ of Celgene Corporation on Monday is shown as a function of elapsed time $t-t_0$ since market opening $t_0$. The inset with a zoom into the first minutes demonstrates bursty changes of the spread.  Sometimes the spread stays constant for almost a minute, then it fluctuates strongly on much shorter time scales. During the whole trading day with a total length of 390 minutes the spread performs strong fluctuations between the smallest possible value of one cent and larger values. To see the trend of the spread during the day, the data is also repeated on a logarithmic scale. The curve begins directly after the first change of the spread during the trading hours. Hence, the last spread of the pre-market hours is ignored. 
For calculating the moving average, we divide the trading hours into time windows of varying size and calculate for every time window the arithmetic mean. The first two intervals each span 64 spread changes. Each following interval spans two times more spread changes than its predecessor (the $n$-th interval with $n>2$ spans $64 \cdot 2^{n-2}$ spread changes). The variable duration of time intervals helps adapting to the logarithmic scale and to largely varying bursty activity in different stocks. The moving average illuminates that the spread follows a decreasing trend throughout the whole trading day. Further, the moving average stays far above the smallest possible value of one cent. We are dealing with a so called small-tick stock, because the tick size of one cent is relatively small compared with the spread. 

\begin{figure}[htb]
\begin{center}
    \includegraphics[width=1.0\columnwidth]{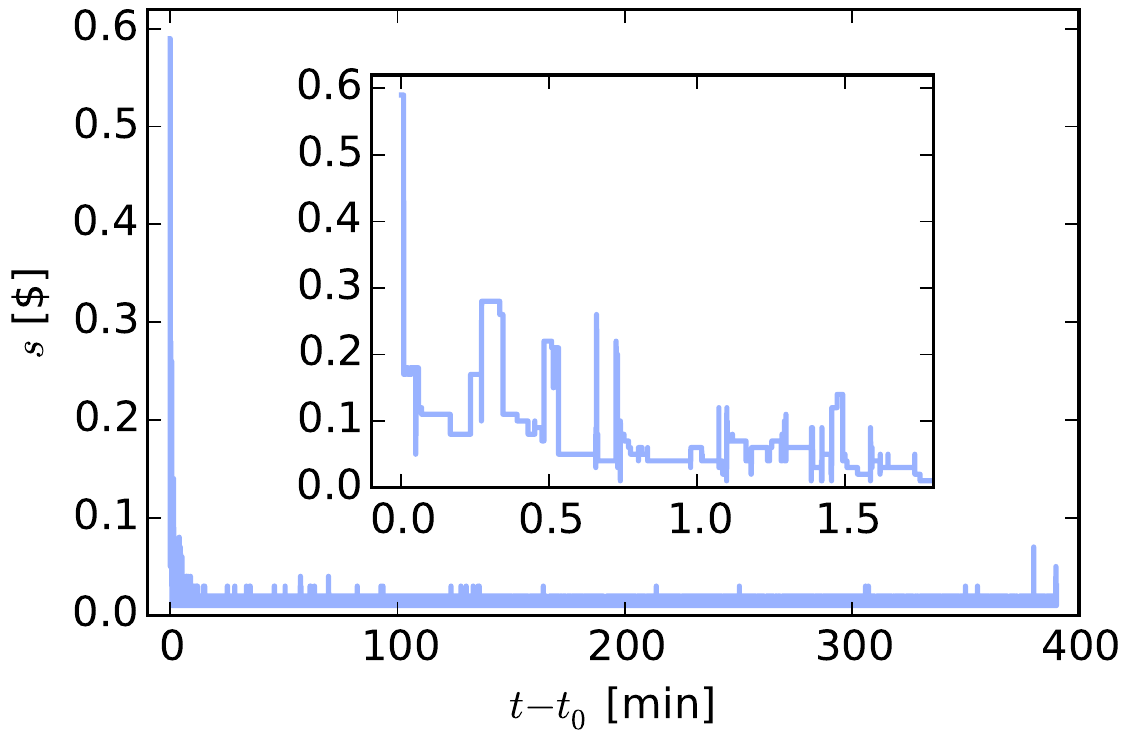}
    \includegraphics[width=1.0\columnwidth]{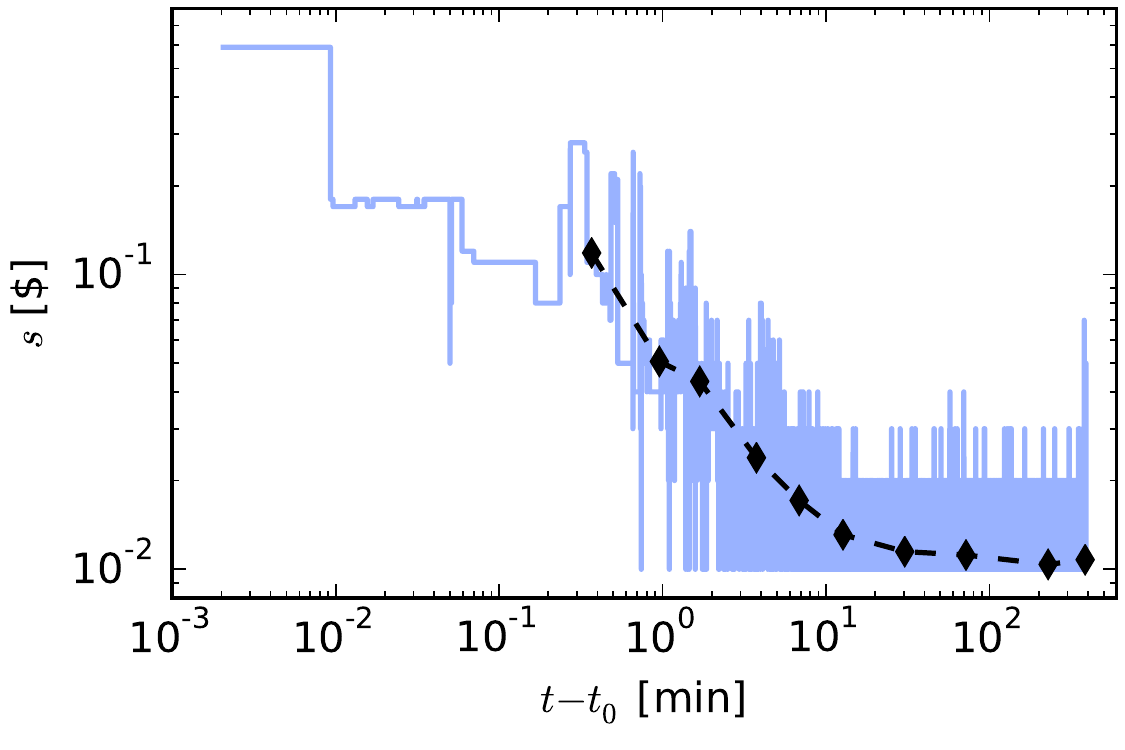}
    \caption{As in Fig.~\ref{fig:spread_CMCSA} for Comcast Corporation, where the spread at later times approaches the smallest possible value of one cent.}
    \label{fig:spread_CMCSA}
\end{center}
\end{figure}

The spread of some other stocks closes so much that the tick size becomes a relevant lower bound, therefore they are called large-tick stocks. As an example Fig.~\ref{fig:spread_CMCSA} shows the 
spread of the stock of Comcast Corporation. The moving average approaches the smallest possible value of one cent after about ten minutes. 
\begin{figure}[htb]
\begin{center}
    \includegraphics[width=1.0\columnwidth]{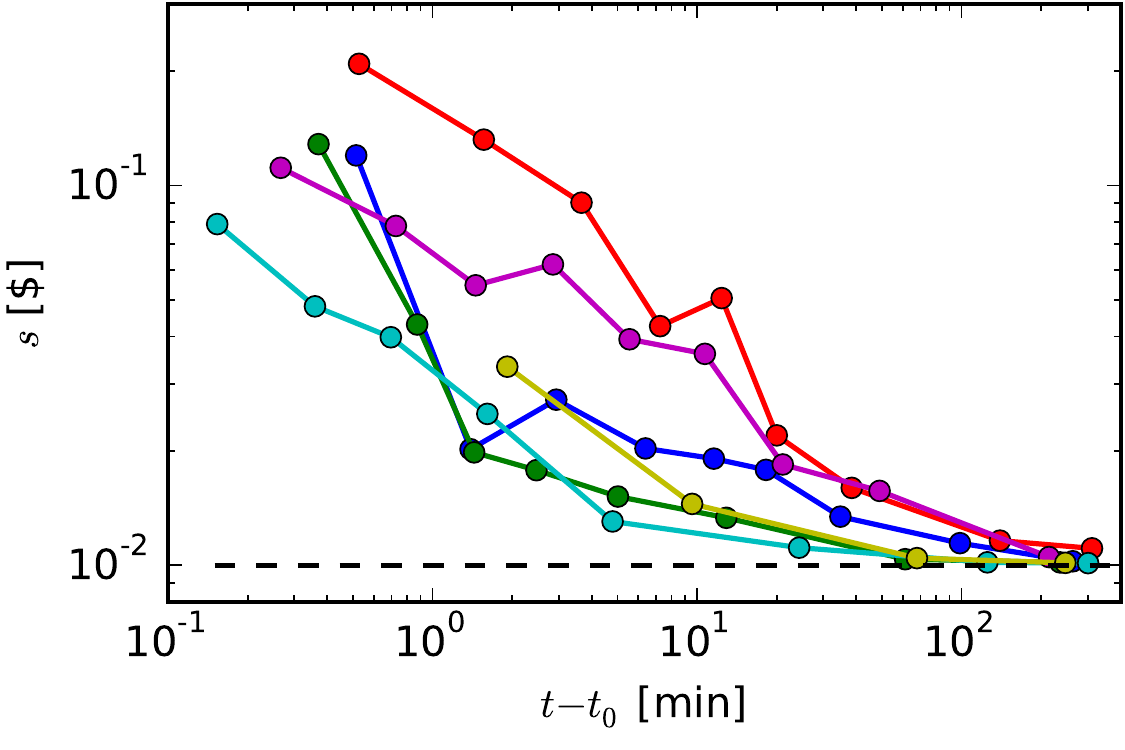}
    \includegraphics[width=1.0\columnwidth]{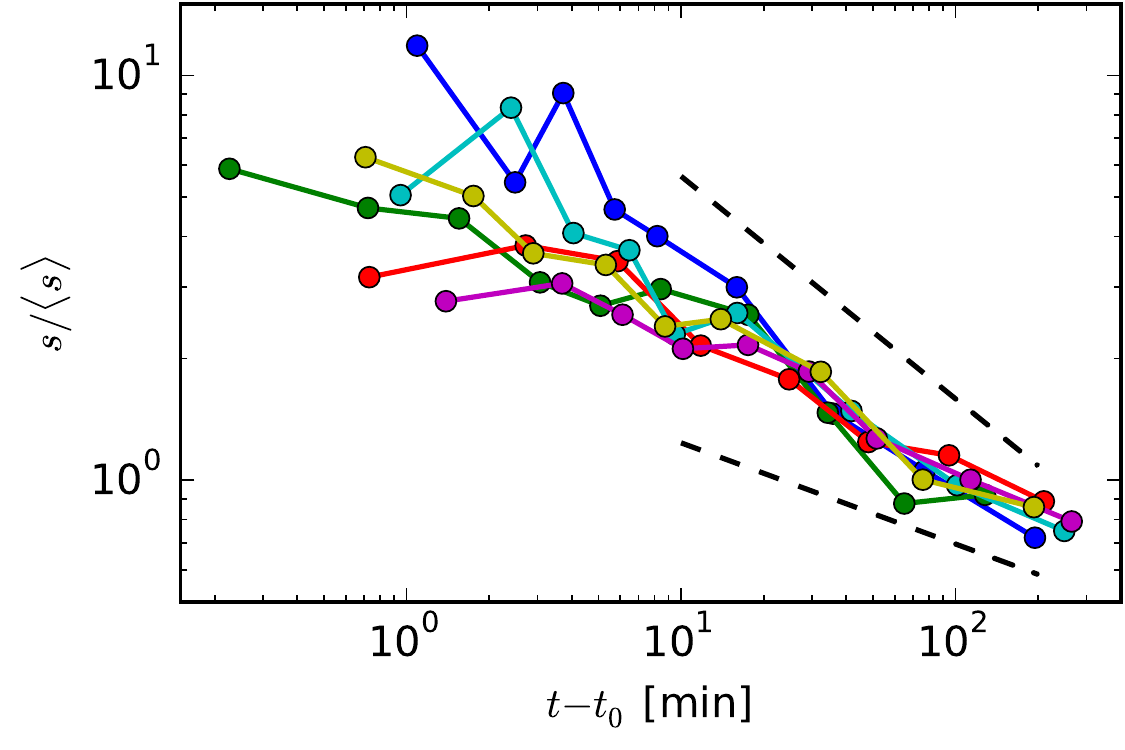}
    \caption{Upper panel: Moving average of the spread for large-tick stocks on double-logarithmic scales (Activision Blizzard in blue, Ebay in green, Liberty Global PLC Class A in red, Micron Technology in cyan, Starbucks in magenta and VIP Industries in yellow). The black dashed line indicates the smallest possible value of one cent. Lower panel: Moving average of the normalized spread for small-tick stocks on double logarithmic scaling (Tripadvisor in blue, Netflix in green, Henry Schein in red, Dollar Tree in cyan, Biogen in magenta and Adobe Systems in yellow. The average of the last time window is omitted, as it can contain only a few spread changes). The black dashed lines are power laws $\propto(t-t_0)^{-\alpha}$ with exponents $\alpha=0.4\pm0.15$.}
    \label{fig:spread_scaling}
\end{center}
\end{figure}
The upper panel of Fig.~\ref{fig:spread_scaling} displays moving averages of the spread on Tuesday for different large-tick stocks. By comparing with the dashed line at one cent, we see that all stocks converge to one tick during the trading day. This behavior is observed for 30 out of the 96 stocks under consideration. We marked the tickers of these stocks with a star in Tab.~1 in the Supplementary Material. The spread of the large-tick stock with ticker SIRI (not listed in the table) is special, because it closes so fast, that the first value of the moving average is already very close one tick. 

The lower panel of Fig.~\ref{fig:spread_scaling} displays moving averages of the normalized spread $s(t)/\langle s\rangle$ for different small-tick stocks on Tuesday.  Here, $\langle s\rangle$ denotes the average of the spread over the whole trading day. After a highly fluctuating opening period of about 20 minutes, the curves start to collapse after around 10 minutes with a scaling behavior $\propto (t-t_0)^{-\alpha}$ for the rest of the trading day. Compared with the dashed lines with a scaling 
\begin{align}
s \propto (t-t_0)^{-0.4\pm0.15},
\end{align}
the exponent $\alpha$ is around $0.4$ for all displayed stocks. All 66 small-tick stocks behave like this. 

This finding can be understood as emergence of a universal law for the intra-day seasonality of stock spreads. While sparse actions of traders directly after market opening imply an erratic behavior with large fluctuations, accumulating more and more actions over time helps the market to converge to a non-equilibrium non-stationary emergent state with stylized time dependence. Only if the spread reaches its technical lower bound of one tick it stays constant for the rest of the trading day, implying a quite different kind of market situation. 

For a better understanding of the rescaling with $\langle s\rangle$, it is worth discussing an idealized spread 
\begin{align}
s^{(\rm id)}(t-t_0)&=c (t-t_0)^{-\alpha}
\end{align}
with $0<\alpha<1$. Because of the pole at $t=t_0$ we define $s^{(\rm id)}$ for times $t$ with $t_0<t_l\leq t\leq t_e$ where $t_e$ marks the end of the trading day. The ratio  $s^{(\rm id)}(t_e-t_0)/\langle s^{(\rm id)}\rangle$ describes how much the spread at the end of the trading day lies below the average. With 
\begin{align}
\langle s^{(\rm id)}\rangle &= \frac{1}{t_e-t_l}\int\limits_{t_l}^{t_e} c (t-t_0)^{-\alpha} {\rm d}t  
\approx  \frac{s^{(\rm id)}(t_e-t_0)}{1-\alpha}, 
\end{align}
where we use $t_l-t_0\ll t_e-t_0$, we find 
\begin{align}
s^{(\rm id)}(t_e-t_0)/\langle s^{(\rm id)}\rangle\approx (1-\alpha).\label{eq:s_final}
\end{align}
Projecting the rescaled moving average of the spread in the lower panel of  Fig.~\ref{fig:spread_scaling} to $t_e-t_0=390\,$min, we find a value around $0.6$. This result is compatible with our finding for an idealized spread with $\alpha=0.4$. For the spread at the starting time $t_l$ shortly after market opening we find $s^{(\rm id)}(t_l-t_0)/\langle s^{(\rm id)}\rangle\approx (1-\alpha) [(t_e-t_0)/(t_l-t_0)]^{\alpha}$. Due to the pole of $s^{(\rm id)}$, the idealized spread at the market opening depends strongly on $t_l$. 
Furthermore, as there are large fluctuations of real data $s(t)$ for the spread in the period after market opening, the comparison with the idealized curve of $s^{(\rm id)}(t)$ at this early elapsed time yields limited information. The scale free time dependence of the spread $s^{(\rm id)}$ also makes it difficult to determine a typical time $t_s$ for the closing of the spread, because of the scaling
\begin{align}
s^{(\rm id)}(t-t_0)=c_1 \left(\frac{t-t_0}{t_s-t_0}\right)^{-\alpha}
=c_2 \left[\frac{t-t_0}{k (t_s-t_0)}\right]^{-\alpha}
\end{align}
with any positive $k$, making the choice of $t_s$ arbitrary. 

\section{Duration of the market opening period}\label{sec:identification}

Our findings so far imply that the spread converges either to a slowly decaying power law, or to a constant value of one tick, if it reaches this technical lower bound. In both cases the convergence takes place during a strongly fluctuating opening period. 
We now introduce a method for identifying the time $t_1$ which marks the end of this opening period, based on the moving average. We calculate the moving average in a modified way with increased time resolution. As before, the first two non-overlapping time windows each span 64 spread changes. The third window overlaps the second window, as it starts at spread change number $1.1\cdot 64$ (the closest integer) and ends at spread change number $1.1\cdot 128$. The next windows are found in the same way, by increasing the number of spread changes with a factor of $1.1$ in each step. The moving average is called $\bar{s}(t)$.
The time of the end of the opening period $t_1$ is defined as the largest time $t$ for which $\bar{s}(t)>3\langle s\rangle /2$. The market opening duration is 
\begin{align}
T &=t_1-t_0,
\end{align}
according to this interpretation. 

\begin{figure}[htb]
\begin{center}
    \includegraphics[width=1.0\columnwidth]{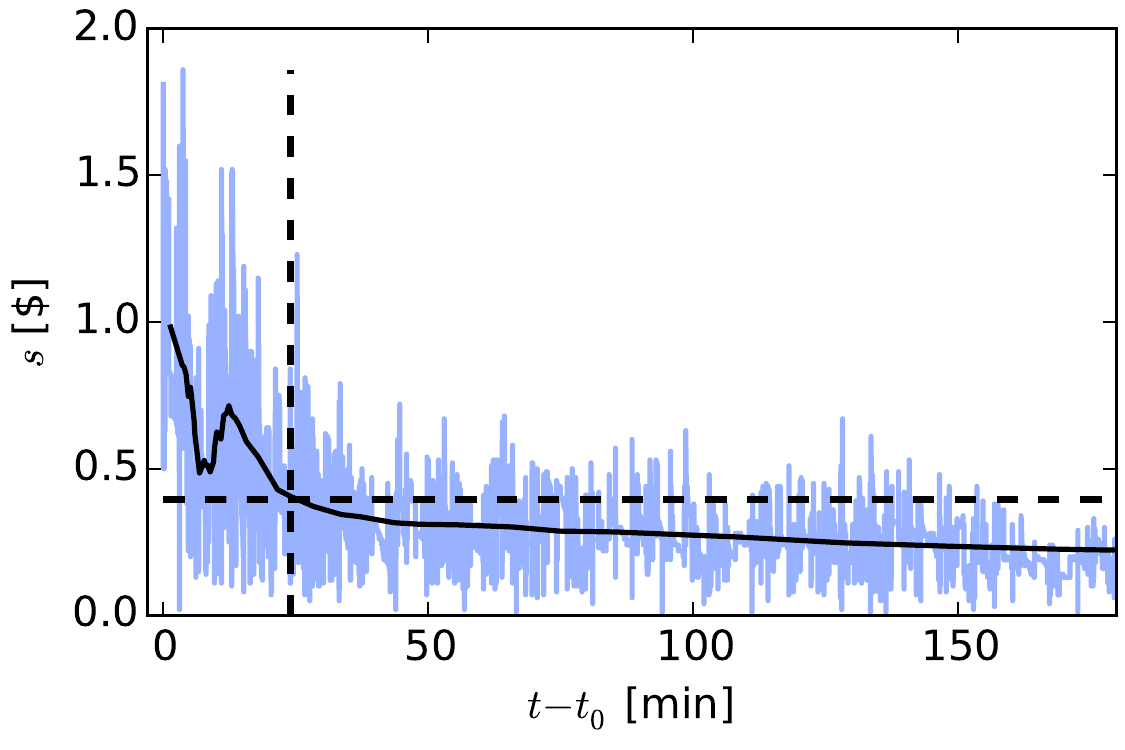}
    \caption{Spread of Illumina Inc. on Monday in blue. The black solid curve illustrates the moving average and the dashed horizontal line is at 3/2 of the average spread. The vertical dashed line highlights the time $t_1$.}
    \label{fig:opening_ILMN}
\end{center}
\end{figure}

This procedure is illustrated in Fig.~\ref{fig:opening_ILMN} for Illumina Inc. on Monday. The strong fluctuations of the spread are reduced in the moving average. The moving average ends before the trading day ends, because the largest possible value of the upper limit of the time window is the end of the trading day. If this value is reached, we allow the lower limit to further increas, as long as the time window spans at least one quarter of all spread changes. The horizontal black line marks the value of $3\langle s\rangle/2$, and the vertical line marks $t_1$, where the $\bar{s}$ is above $3\langle s\rangle/2$ for the last time. We see that the spread still changes considerably after this time, but slowly. 

\begin{figure}[htb]
\begin{center}
    \includegraphics[width=1.0\columnwidth]{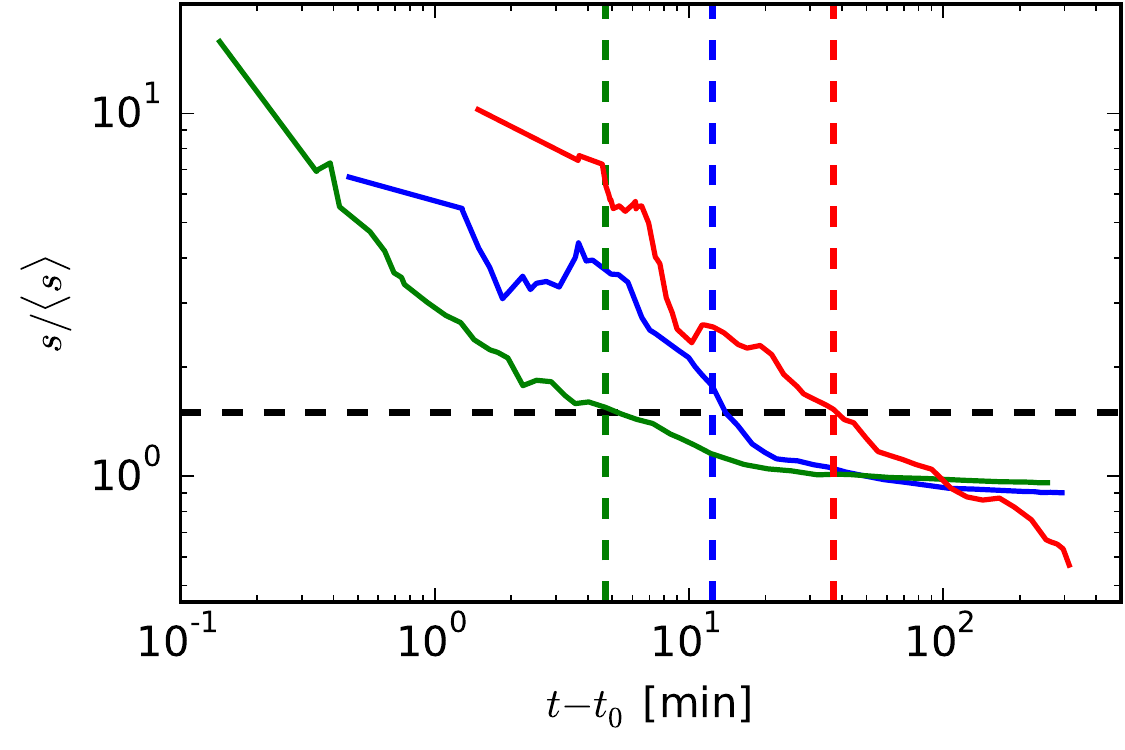}
    \includegraphics[width=1.0\columnwidth]{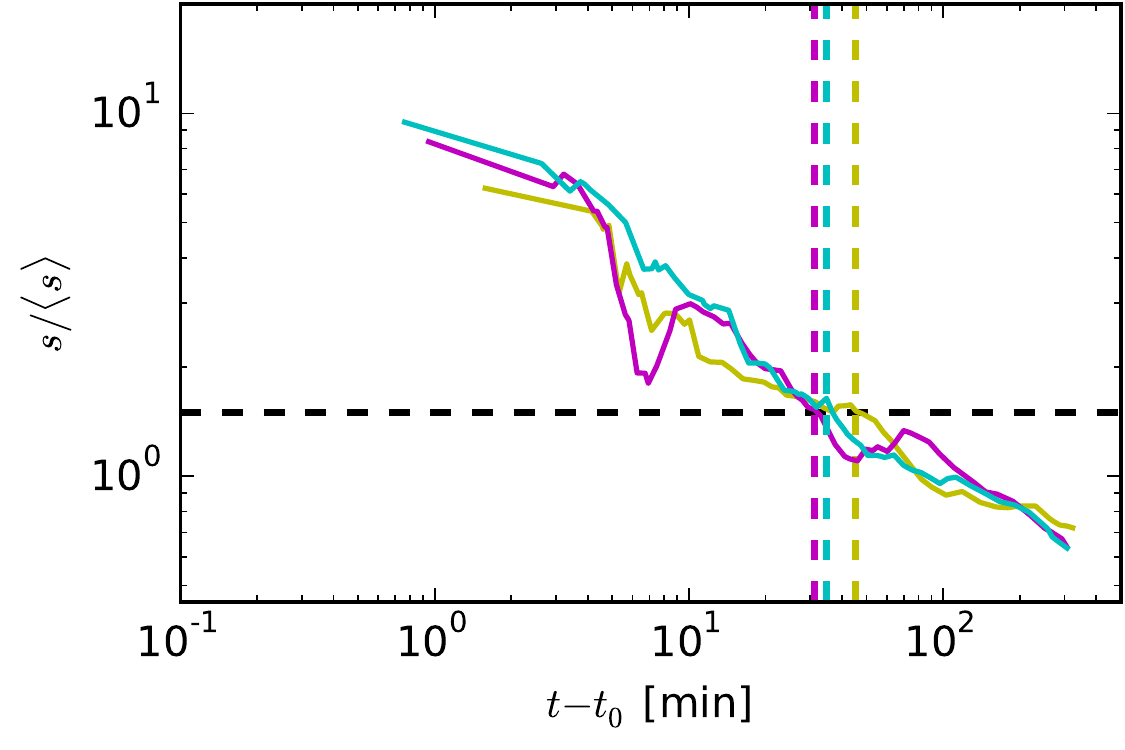}
    \caption{Upper panel: Rescaled moving average of the spread on Friday on a double logarithmic scale (Activion Blizard in green, Yahoo in blue, Autodesk in red). Vertical dashed lines mark the end of the market opening period $t_1$, the horizontal dashed line is at 3/2. Lower panel: Rescaled moving average of the spread of Lam Research Corporation (Monday in yellow, Wednesday in magenta, Friday in cyan).}
    \label{fig:opening_various}
\end{center}
\end{figure}

In the upper panel of Fig.~\ref{fig:opening_various} the rescaled moving average of the spread $\bar{s}/\langle s\rangle$ is shown for Activion Blizard, Yahoo and Autodesk. For Activition Blizard and Yahoo the rescaled spread saturates to a value slightly below one. These are large-tick stocks, their spread saturates to one tick at the end of the trading day. The vertical dashed lines indicate short opening durations $T=t_1-t_0$ of less than five minutes for Activation Blizard and about twelve minutes for Yahoo. Afterwards, the spread only changes by a small factor below two and saturates fast. The reascaled spread of Autodesk behaves differently with a late $t_1$ about 37 minutes after market opening and a decreasing spread during the whole rest of the trading day. Here the moving average of the spread reduces by more than a factor of two after $t_1$ which can be understood with Eq.~(\ref{eq:s_final}). The longer lasting opening period and the substantial change of the spread during the rest of the trading day is typical for small-tick stocks with a large spread. Comparing the three stocks, the opening duration varies by a factor of eight. In the lower panel of Fig.~\ref{fig:opening_various} data is provided for Lam Research Corporation on three different trading days (Monday, Wednesday and Friday). All curves are similar, and accordingly the times $t_1$ vary by less than a factor $3/2$ between 31 minutes and 45 minutes. The largest differences between the curves appear in the opening period due to large fluctuations.

\section{Comparison across stocks}\label{sec:opening}

We proceed by calculating market opening durations $T=t_1-t_0$ for all stocks,  except for Sirius XM Holdings Inc (ticker SIRI) as the spread of this stock starts close to one tick and has a flat evolution during the trading day. By ignoring this stock, we are left with 95 stocks for all further investigations. All stocks are listed in Tab.~1 in the Supplementary Material. 

\begin{figure}[htb]
\begin{center}
    \includegraphics[width=1.0\columnwidth]{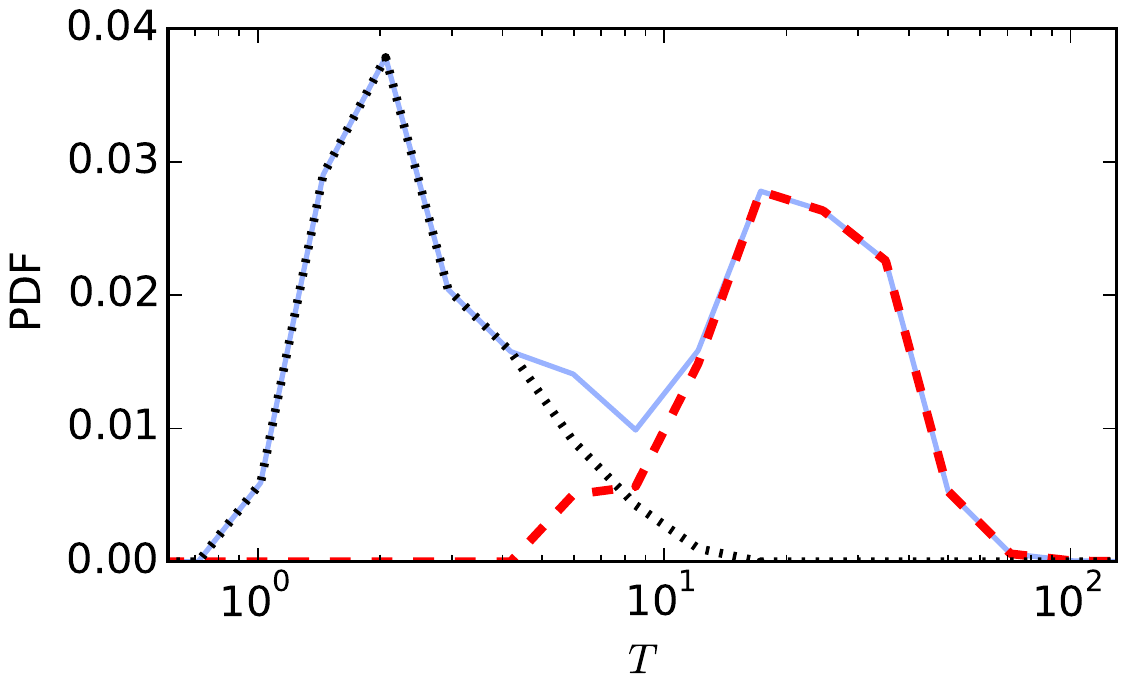}
    \includegraphics[width=1.0\columnwidth]{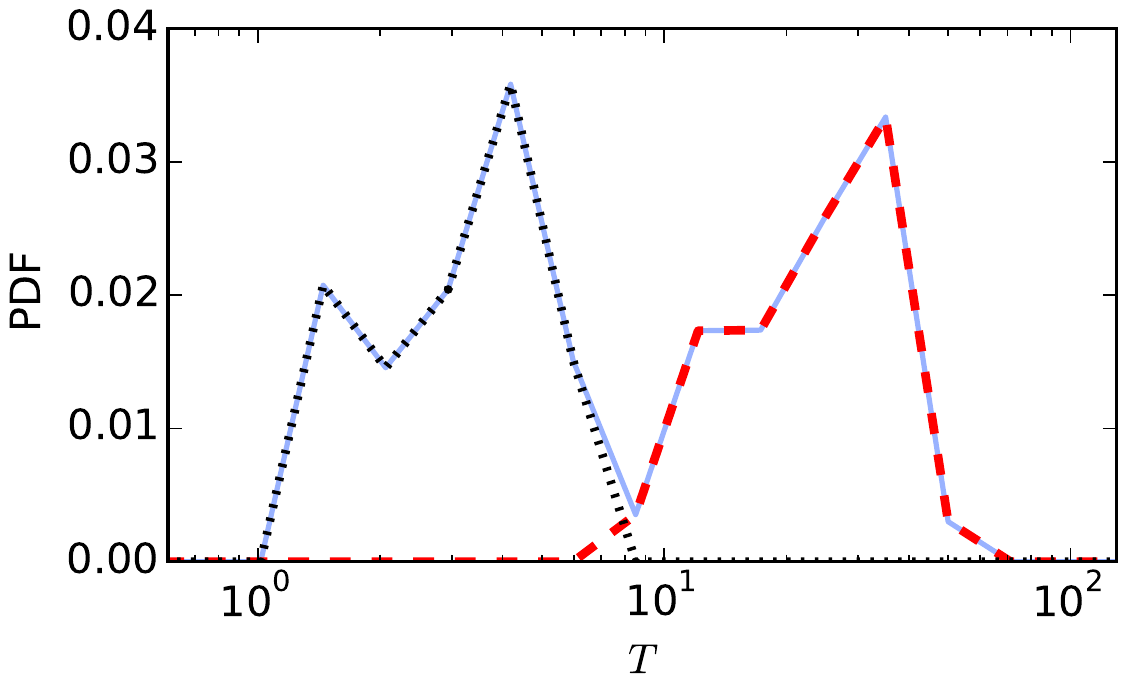}
    \caption{Upper panel: PDF of market opening durations on a logarithmic scale with the solid line. The fraction of opening durations connected to large-tick stocks with smallest spread is highlighted with the dotted black line, the fraction connected to all other stocks with the red dashed line. Lower panel: Same for opening durations averaged over five trading days for every stock individually.}
    \label{fig:opening_pdf}
\end{center}
\end{figure}

In the upper panel of Fig.~\ref{fig:opening_pdf} we display the probability density function (PDF) of the opening durations $T$ for all considered stocks and weekdays. Opening durations span a surprisingly broad range from about a minute to more than one hour. We conclude that the opening phases are very different for different stocks. 
The dotted black line indicates the fraction of the PDF originating from large-tick stocks with a particularly small average spread ($\langle s\rangle<0.0115\,\$$ with the spread being averaged over all five trading days). These 12 stocks (tickers AMAT, CSCO, EBAY, FOXA, INTC, MSFT, MU, SPLS, SYMC, VIP, VOD and YHOO) lead to small opening durations. This result can be understood from Fig.~\ref{fig:opening_various}: For such stocks the opening period is short because the spread saturates fast to a value close to one tick. Still, the stocks differ with respect to the time scale, on which the saturation takes place. 
With the red dashed line in Fig.~\ref{fig:opening_pdf} we learn that the 83 stocks with larger spread have larger opening durations. The area under the red dashed curve is more than six times larger than the area under the black dotted curve. The different optical appearance originates from logarithmic scaling. 

Averaged over the five working days, Fig.~\ref{fig:opening_pdf} (lower panel) shows that even averaged durations span a large range of values, and the distinction between large-tick stocks and small-tick stocks stays the same. Averaged durations for individual stocks are listed in Tab.~1 in the Supplementary Material.

\section{Comparison with the rest of the trading day}\label{sec:comparison}

We now analyze, how the order flow during the market opening period differs from the order flow during the rest of the trading day. For a given stock and trading day, we use the opening duration $T$ and define three time intervals, each of duration $T$. The first interval spans the opening period from $t_0=$ 9:30 am to $t_1=t_0+T$. The second interval starts at noon. It spans the time interval from $t_2=$ 12 am to $t_3=t_2+T$. The third interval spans the time from $t_4=$ 2 pm to $t_5=t_4+T$. 
As a price volatility measure, we consider the standard deviation $\sigma$ of one minute returns 
\begin{align}
\sigma &= \sqrt{\langle r_{1\,\rm min}^2\rangle-\langle r_{1\,\rm min}\rangle^2},\\
r_{1\,\rm min} &= \frac{m(t+1\,{\rm min})-m(t)}{m(t)},
\end{align}
with the time average $\langle r_{1\,\rm min} \rangle$ ranging over the time interval under consideration. Based on the values $\sigma$ for all stocks, trading days and time intervals, we compute separate PDFs for the opening period, the time interval starting at noon and the time interval starting at 2 pm. 
\begin{figure}[htb]
\begin{center}
    \includegraphics[width=1.0\columnwidth]{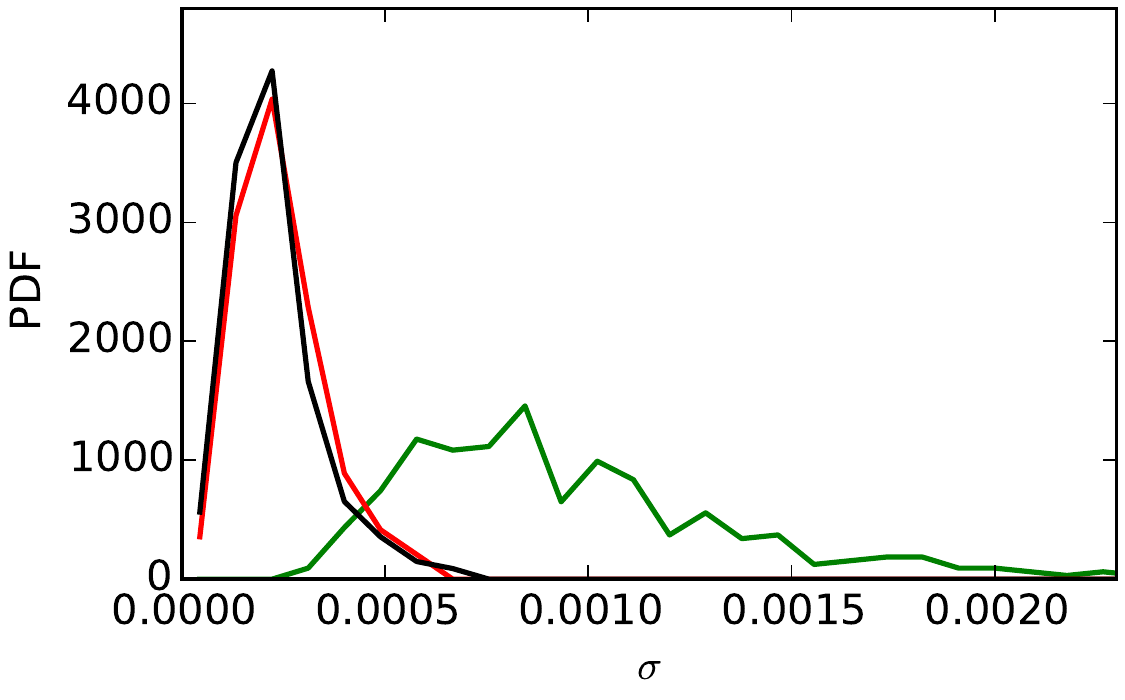}
    \caption{PDF of one minute return standard deviation during the opening (green), after noon (red) and after 2:00 pm (black).}
    \label{fig:price}
\end{center}
\end{figure}
As Fig.~\ref{fig:price} displays, the price volatility is much larger for the opening period than for the later time intervals. 
A possible explanation are news which arrive over night and have to be included in the price during the opening period. 

\begin{figure}[htb]
\begin{center}
    \includegraphics[width=1.0\columnwidth]{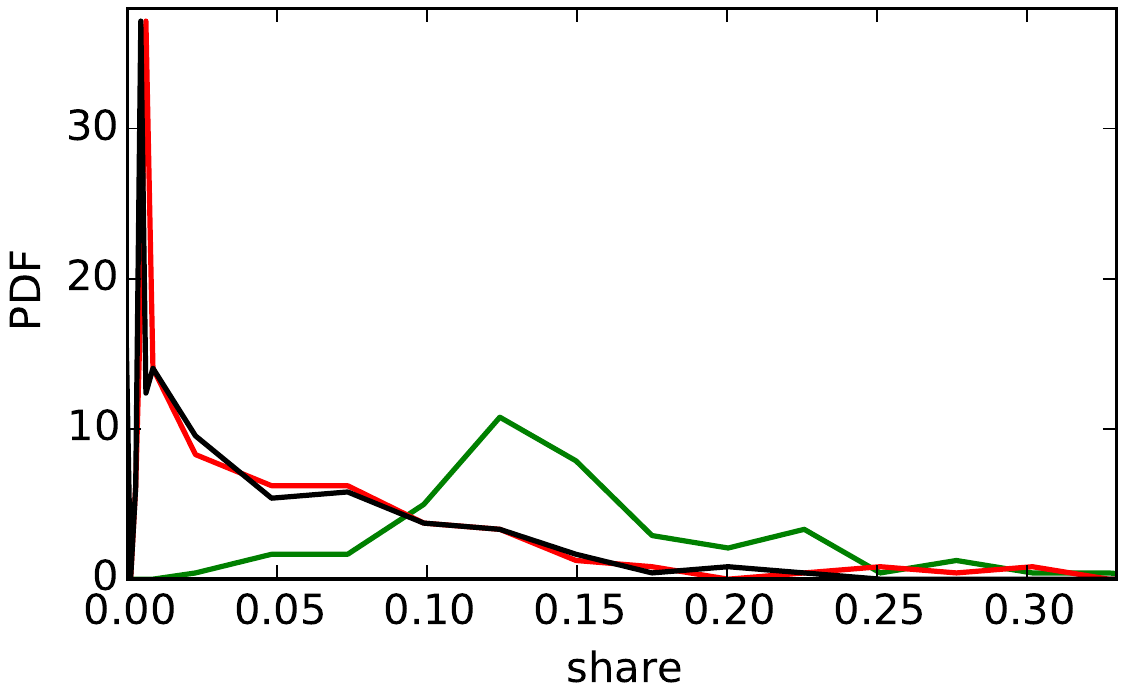}
    \caption{PDF of the share of in-spread limit orders among all limit orders during the opening (green), after noon (red) and after 2:00 pm (black).}
    \label{fig:share}
\end{center}
\end{figure}

As can be inferred from Fig.~\ref{fig:share}, the order flow during the opening period differs from the rest of the day. We calculate for every stock, weekday and time interval the share of in-spread limit orders among all limit orders. In-spread limit orders are limit orders that are placed between best ask price and best bid price and therefore change the quote. For the opening period, we calculate the PDF of in-spread limit order shares across all stocks and weekdays. Analog results are shown for the second and third time window. The trading at opening has larger shares of limit orders that change the quotes than during the rest of the trading day.

\section{Discussion}\label{sec:summary}

We compared properly smoothed and rescaled time series of the spread of stocks traded on NASDAQ. 
We found that the data collapses to a power law with small negative exponent for the majority of the stocks, namely all 66 large-tick stocks. This means that the spread is non-stationary during the whole trading day and closes slowly. We conclude that the market converges to a non-stationary emergent state which follows a stylized time dependence. 

Only if the spread reaches the technical lower bound of one tick, it stays constant for the rest of the trading day. This behavior was found for a minority of 30 small-tick stocks. As was known already from other studies, the order flow of small-tick stocks and large-tick stocks has major differences~\cite{gareche2013fokker,dayri2015large}. Here we found a complementary aspect of this fact. The technical constraint keeps the market for large-tick stocks from further adjusting the spread during the trading day, which would be the natural non-stationary emergent behavior otherwise. 

Our findings have implications for data analysis of the order flow. If the market opening is excluded from the data analysis, this should be done individually for every stock. There are also implications on agent based modeling. The order flow should be modeled with a distinct opening phase, during which the order book is filled and the spread stabilizes. 
Further the non-stationarity during the rest of the trading day has to be accounted for. 
These results have to be included when setting up agent-based models for stock markets. 

We further elaborated on the highly fluctuating opening period during which the spread is large and the scaling behavior emerges. The end of this period was identified as the time where the smoothed spread falls below two thirds of its average. 
Surprisingly, we found that opening durations vary from a few minutes to more than one hour. 
The largest differences appear between large-tick stocks with particularly small spread, which reach the lower bound of one tick very quickly, and small-tick stocks. 
In a second step, we used the individual durations for characterizing the order flow  during the opening period across the whole market. We compared with later trading periods around noon and in the early afternoon. The price and order flow behaves distinctly different during the opening period than later. The volatility is much larger, and the share of limit orders being placed between best buy and best sell offer is much larger during the opening period. 
In Summary, these results give a first example of how stylized facts emerge during the opening of stock markets. 

\bibliography{bibli}

\pagebreak
\onecolumngrid

\section*{Supplementary marerial to \\ \textit{Emergence of stylized facts during the opening of stock markets}}

{
\centering 

Sebastian M.~Krause, Jonas A. Fiegen and Thomas Guhr

}
\vspace{18mm}

\begin{table}[htb]
{
\small
\centering
\begin{minipage}{0.31\columnwidth}
  \begin{tabular}{c c c}
  Ticker & $T\,$[min] & $Q$\\
  \hline
AAL* & 19.7 & 1.2 \\
AAPL* & 10.0 & 2.2 \\
ADBE & 27.4 & 2.2 \\
ADI & 28.1 & 1.5 \\
ADP & 32.8 & 2.3 \\
ADSK & 32.7 & 1.4 \\
AKAM & 35.0 & 2.1 \\
ALXN & 25.2 & 2.2 \\
AMAT* & 3.1 & 3.5 \\
AMGN & 36.1 & 2.4 \\
AMZN & 21.5 & 2.0 \\
ATVI* & 14.4 & 1.5 \\
AVGO & 39.0 & 1.9 \\
BBBY & 24.1 & 2.1 \\
BIDU & 28.3 & 2.6 \\
BIIB & 32.6 & 1.7 \\
BMRN & 29.4 & 1.7 \\
CA* & 12.7 & 1.4 \\
CELG & 36.5 & 1.7 \\
CERN & 31.9 & 5.1 \\
CHKP & 31.9 & 1.7 \\
CHRW & 30.5 & 2.9 \\
CHTR & 34.4 & 1.7 \\
CMCSA* & 10.3 & 2.0 \\
COST & 30.1 & 2.1 \\
CSCO* & 1.3 & 5.8 \\
CTSH & 22.8 & 1.3 \\
CTXS & 42.0 & 1.7 \\
DISCA* & 17.8 & 1.3 \\
DISH & 28.8 & 1.7 \\
DLTR & 33.6 & 1.7 \\
EA & 30.7 & 2.0 
  \end{tabular}
  \end{minipage}
  \begin{minipage}{0.31\columnwidth}
  \begin{tabular}{c c c}
  Ticker & $T\,$[min] & $Q$\\
  \hline
EBAY* & 4.1 & 3.8 \\
EQIX & 27.4 & 2.5 \\
ESRX & 28.1 & 3.1 \\
EXPD & 35.7 & 2.6 \\
FAST & 42.6 & 2.2 \\
FB & 19.2 & 2.0 \\
FISV & 37.4 & 2.0 \\
FOXA* & 6.7 & 5.1 \\
GILD & 23.9 & 2.5 \\
GOOG & 22.1 & 3.0 \\
GRMN & 33.7 & 2.3 \\
HSIC & 34.7 & 4.1 \\
ILMN & 24.4 & 2.3 \\
INTC* & 2.5 & 2.3 \\
INTU & 32.0 & 2.6 \\
ISRG & 21.4 & 1.9 \\
JD* & 10.9 & 2.7 \\
KHC & 42.2 & 4.4 \\
KLAC* & 28.3 & 1.7 \\
LBTYA* & 15.0 & 1.9 \\
LLTC & 19.4 & 1.4 \\
LMCA & 34.5 & 2.0 \\
LRCX & 34.3 & 1.7 \\
LVNTA & 34.7 & 1.8 \\
MAR & 31.1 & 2.3 \\
MAT* & 14.6 & 1.2 \\
MDLZ* & 14.0 & 2.4 \\
MNST & 32.9 & 2.1 \\
MSFT* & 4.0 & 4.8 \\
MU* & 4.3 & 4.5 \\
MYL & 20.9 & 1.9 \\
NFLX & 30.9 & 1.2 
  \end{tabular}
  \end{minipage}
  \begin{minipage}{0.31\columnwidth}
  \begin{tabular}{c c c}
  Ticker & $T\,$[min] & $Q$\\
  \hline
NTAP* & 13.5 & 2.5 \\
NVDA* & 12.9 & 1.6 \\
NXPI & 35.4 & 2.8 \\
ORLY & 49.0 & 1.7 \\
PAYX & 35.5 & 2.5 \\
PCAR & 35.8 & 1.7 \\
PCLN & 35.2 & 10.2 \\
QCOM* & 13.9 & 2.4 \\
REGN & 19.7 & 3.9 \\
ROST & 28.3 & 1.6 \\
SBAC & 31.4 & 1.9 \\
SBUX* & 16.8 & 1.4 \\
SNDK & 22.4 & 2.0 \\
SPLS* & 4.9 & 4.1 \\
SRCL & 38.6 & 1.8 \\
STX & 23.4 & 1.8 \\
SYMC* & 5.3 & 4.7 \\
TRIP & 35.4 & 1.6 \\
TSCO & 34.7 & 1.7 \\
TSLA & 37.8 & 6.7 \\
TXN* & 18.9 & 1.3 \\
VIAB & 30.7 & 2.2 \\
VIP* & 3.8 & 3.2 \\
VOD* & 2.3 & 4.2 \\
VRSK & 40.2 & 2.0 \\
VRTX & 35.8 & 1.7 \\
WDC & 30.2 & 1.4 \\
WFM* & 23.0 & 1.8 \\
WYNN & 41.9 & 1.5 \\
XLNX & 24.7 & 2.5 \\
YHOO* & 6.6 & 5.0 \\
 & &
  \end{tabular}
  \end{minipage}
}
  \caption{Stocks analyzed. Tickers of large-tick stocks, where the moving average of the spread approaches one tick, are marked with a star. Opening period duration $T$ averaged over five trading days and quotient $Q$ between largest opening duration and smallest opening duration of a stock.}\label{tab:durations}
  \end{table}

  
In Tab.~\ref{tab:durations} the opening durations $T$ averaged for every stock over all five working days are listed for all considered stocks. The quotient $Q$ between the maximal opening duration and the minimal opening duration is also listed. For Priceline Group Inc. (ticker PCLN) this quotient is larger than ten, meaning that opening durations are very diverse for this stock. For most other stocks, the quotient $Q$ is much smaller.

\end{document}